Development of the algorithm for differentiating bone metastases and trauma of the ribs in bone scintigraphy and demonstration of visual evidence of the algorithm     -Using only anterior bone scan view of thorax-

Shigeaki Higashiyama 1
Yukino Ohta 2
Yutaka Katayama 3
Atsushi Yoshida 1
Joji Kawabe 1

1 Department of Nuclear Medicine, Graduate School of Medicine, Osaka City University
2 MedCity21, Division of Premier Preventive Medicine, Osaka City University Hospital
3 Department of Radiology, Osaka City University Hospital

Corresponding Author
Shigeaki Higashiyama
https://orcid.org/0000-0002-3976-4521
Department of Nuclear Medicine, Graduate School of Medicine, Osaka City University
1-4-3, Asahimachi, Abeno-ku,Osaka 545-8585

Abstract

Background:

Although there are many studies on the application of artificial intelligence (AI) models to medical imaging, there is no report of an AI model that determines the accumulation of ribs in bone metastases and trauma only using the anterior image of thorax of bone scintigraphy. In recent years, a method for visualizing diagnostic grounds called Gradient-weighted Class Activation Mapping (Grad-CAM) has been proposed in the area of diagnostic images using Deep Convolutional Neural Network (DCNN). As far as we have investigated, there are no reports of visualization of the diagnostic basis in bone scintigraphy. Our aim is to visualize the area of interest of DCNN, in addition to developing an algorithm to classify and diagnose whether RI accumulation on the ribs is bone metastasis or trauma using only anterior bone scan view of thorax.

Material and Methods:

For this retrospective study, we used 838 patients who underwent bone scintigraphy to search for bone metastases at our institution. A frontal chest image of bone scintigraphy was used to create the algorithm. We used 437 cases with bone metastases on the ribs and 401 cases with abnormal RI accumulation due to trauma.

Result:

AI model was able to detect bone metastasis lesion with a sensitivity of 90.00% and accuracy of 86.5%. And it was possible to visualize the part that the AI model focused on with Grad-CAM.

1. Introduction

Bone scintigraphy is the most versatile nuclear medicine test for diagnosing metastatic bone lesions [1.2]. When abnormal Radio Isotope (RI) accumulation is observed in the ribs, it is clinically important to distinguish whether the accumulation is due to bone metastasis or post-traumatic accumulation. RI accumulation in bone metastatic lesions extends in the long axis and tangential directions of the ribs [3]. On the other hand, the point of differential diagnosis of the two lesions is that the accumulation of ribs in trauma is punctate accumulation [3]. In order to confirm the accumulation of shapes extending in the tangential direction of the ribs, it is necessary to add an oblique image in addition to the anterior bone scan view of thorax. However, adding an oblique image imaging requires an extension of the examination time. This causes delays in scheduled examinations and is a physical burden on the patient. In facilities where there is no permanent nuclear medicine specialist to determine whether additional imaging of the oblique image is necessary, the problem is that the diagnostic ability is reduced.

For these reasons, it is extremely useful to develop an algorithm for accurately classifying and diagnosing whether abnormal accumulation of bone scintigraphy is bone metastasis or traumatic accumulation by using AI technology only with frontal chest images.

2. Material and Methods

2.1. Study design

The bone scintigraphy image data used in this study was retrospectively collected from Osaka City University Hospital.

The ethics bord of our institution reviewed and approved the protocol of the study in 2019 as ethical No2019-040. The need for informed consent was waived because the images had been acquired during daily clinical practice.

2.2. Study patients

The subject was a case in which bone scintigraphy was performed in search of metastatic bone tumor at Osaka City University Hospital. Among them, 838 cases in which an oblique image was added in addition to the front image for the abnormal accumulation of ribs were included in this study.

We obtained written informed consent from all the participants in accordance with the Code of Ethics of the World Medical Association. All procedures performed in this study were in accordance with the ethical standards of our institutional research committee and with the principles of the 1964 Declaration of Helsinki and its later amendments or comparable ethical standards.

2.3. Image acquisition

Bone scintigraphy is performed after the administration of 740 MBq 99mTc-Hydroxymethylene diphosphonate (HMDP).

The imaging devices employed were an ADAC Forte and Phillips Bright view X equipped with low-energy high-resolution collimators. The imaging parameters of the Forte were a scan time of 250sec, 256 × 256 matrix, and a 140-keV photopeak with a 20% window. Those of the Bright view X were a scan time of 250sec, 256 × 256 matrix, and a 140-keV photopeak with a 20% window.

2.4. Case selection

A frontal view of thorax in bone scintigraphy was used for the development and research of this algorithm. The cases of bone metastasis to the ribs are 437 cases diagnosed as bone metastasis from the image of bone cinch on the oblique image. These cases were diagnosed as bone metastases based on the judgment of either other pixel diagnosis, histological diagnosis by surgery, or follow-up. From the image of bone scintigraphy, it was diagnosed as accumulation in trauma. These cases were diagnosed as having accumulated in trauma due to confirmation of the history of bruising at the site of accumulation of patients and weakening or disappearance of accumulation in follow-up. Two radiologists (11 and 20 years of experience) were in charge of image selection and diagnosis. We divided the cases into training dataset and test dataset. Training dataset contain 347 of metastasis cases and 328 trauma cases, and test dataset contain 90 of metastasis cases and 73 trauma cases.

2.5. AI Model development

To classify the abnormal uptake on bone scintigraphs, we conducted transfer learning which is one of the abilities of the DCNN. In transfer learning, the pre-trained weights that are necessary for calculations in the DCNN are applied as a starting point to learn a new task. Compared with learning from the initial state, the DCNN can learn with a small image in a short time [4]. Moreover, its effectiveness has been reported in various studies [5.6]. In this study, we employed AlexNet architectures [7]. AlexNet contains five convolutional layers used to extract the features identifying targets on the images. In the transfer learning process of this study, only the final layer was replaced to adapt to the target task. We trained the DCNN by using training dataset. The experiments were conducted in the MATLAB 2020a (MathWorks, Natick, MA, USA) environment on Intel® CoreTM i7-9700K CPU (Intel Santa Clara, CA, USA) and GeForce GTX 1060 6GB (NVIDIA Corporation, Santa Clara, CA, USA). The DCNN was trained for 100 epochs with a batch size of 128. We utilized SGDM with a momentum of 0.9, and the learning rate was fixed at 0.0001. To visualize the region in an image on which the DCNN was based to make a decision, we created heat maps, exploiting Class Activation Mapping

(CAM) [8].

To investigate the effect of the number of images on the heat maps and classification accuracies, we trained DCNN by using four datasets: All 675 images, 500 images, 400 images and 300 images. The datasets other than the dataset consisting of all 676 images consisted of the same number of metastatic and traumatic cases. The learning conditions were the same as above.

2.6. Statistical analysis

We evaluated the trained DCNN by using test dataset. Statistical analyses were performed using MATLAB 2020a. The mean of each category was calculated, and the confidence intervals for the means were determined using a normal distribution. To evaluate the performance of the trained DCNN, the classification accuracy, sensitivity, and specificity were calculated. Additionally, receiver operating characteristic (ROC) curves were generated, and the area under the curve (AUC) was determined. In the ROC analysis, the closer the AUC value was to 1, the value shows the better the classification performance. AUC values between 0.5 and 0.7 indicate a poor performance, those between 0.7 and 0.9 indicate a moderate performance, and those between 0.9 and 1 indicate that the tool has high predictability [9].

3. Result

Bone metastasis lesion detection became possible with a sensitivity of 90.00% and accuracy of 86.5%. ROC analysis was able to obtain a diagnostic ability of AUC = 0.9037 (Fig. 1). In addition, it was possible for DCNN to visualize bone metastases and areas of interest (Fig. 2). The left side of Figure 2 is a bone scintigraphy of bone metastases on both upper ribs. The part of the bone metastasis of ribs were depicted in black as an abnormal RI accumulation (arrows). The right side of Figure 2 is a color scale image of the part that Grad-CAM focuses on as a part of bone metastasis. The red part indicates that DCNN is focusing on stronger bone metastases. It can be judged that DCNN is paying attention to the part of bone metastasis.

Additionally, as the number of images in the training dataset increased, the hot spots on the heat maps converged and the classification accuracies improved (Fig.3).

4. Discussion

We developed an AI model for detecting bone metastasis and trauma of rib from only anterior bone scan view of thorax. Our model had a sensitivity (90.00%), specificity (82.19%), accuracy (86.50%), positive predictive value (86.17%), negative predictive value (86.96%) and AUC value (0.9037). In addition to this result, we succeeded in visualizing the part focused on by the AI model with Grad-CAM. To the best of our knowledge, this is the first study to develop detection model for bone scintigraphy only using anterior view of thorax.

Searching for bone metastases before and during treatment for prostate cancer, breast cancer, and lung cancer is indispensable, and bone scintigraphy, which enables diagnosis of systemic bone metastasis with a single injection, is very important [10].

The main finding of abnormal accumulation of bone metastatic lesions is that abnormal accumulation extends in the long axis and tangential directions of the ribs. Additional imaging in the oblique direction is useful for confirming whether the rib accumulation extends in the tangential direction, but the extension of the imaging time is a problem. Diagnosing abnormal RI accumulation on the ribs without adding oblique view is useful in reducing imaging time. The increased accuracy of the algorithm for automatic diagnosis may also be useful for interpreting nuclear medicine examinations in facilities where nuclear medicine specialists are abuse. The creation of our algorithm can shorten the imaging time of bone scintigraphy, which is an appointment test, and it is thought that performing the test without delay will also help increase the number of tests.

It has been reported that the type and performance of DCNN affect the heatmaps [12]. In this study, as the number of images in the training dataset increased, the hot spots on the heat maps converged and the classification accuracies improved. More various training datasets are needed for DCNN to clearly extract the image features required for classification and achieve high classification accuracies.

Conclusions

The AI model we created was able to distinguish between bone metastasis and trauma with high sensitivity and specificity for abnormal accumulation of ribs without additional imaging of the oblique view image. In addition, we visualized the area of interest of the AI model that diagnoses abnormal accumulation of ribs for bone metastases.

Figure legend

Figure 1

The accuracy of DCNN was evaluated by the receiver operating characteristic (ROC) curve. The area under the curve (AUC) was 0.9037.

Figure 2

The image on the left is a bone scintigraphy of bone metastases on both ribs. The image on the right is an image of bone scintigraphy overlaid with a Grad-CAM image. It proves that DCNN is paying attention to the part of bone metastasis.

Figure 3

The figure shows that as the number of training data sets increased, the areas of interest was converged and the classification accuracy improved.

Figure 1

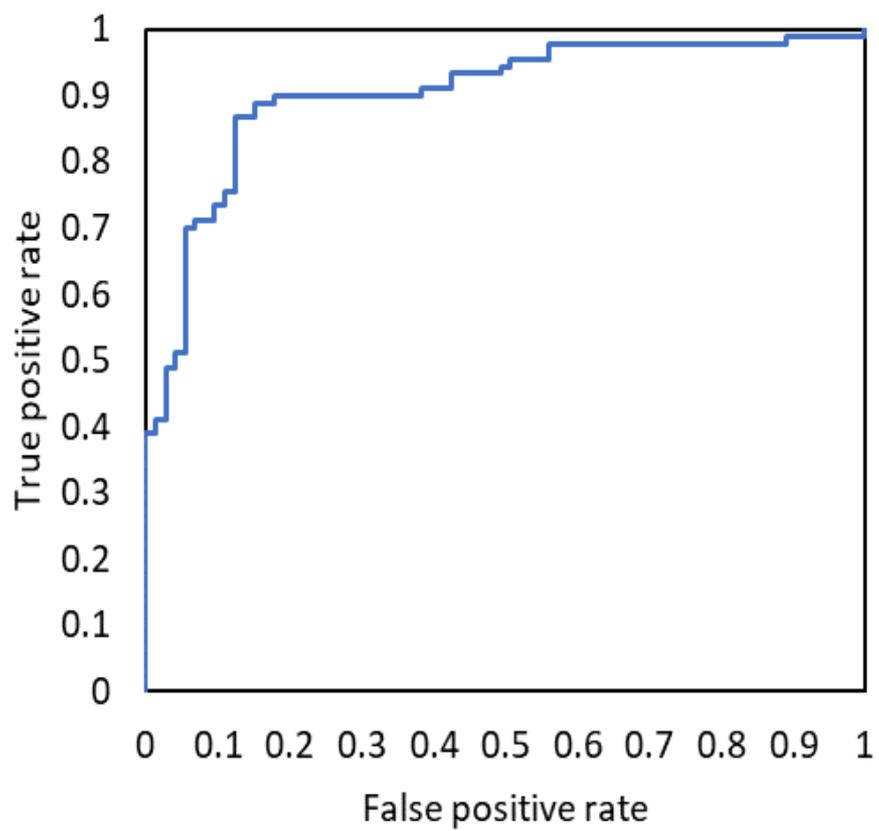

Figure 2

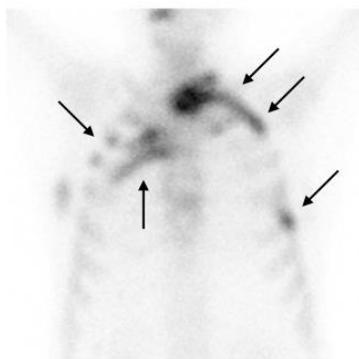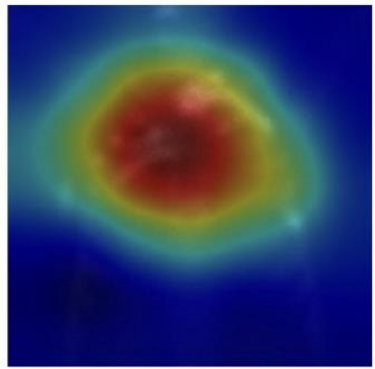

Figure 3

| | Original image | Trainde by using 300 images | Trainde by using 400 images | Trainde by using 500 images | Trainde by using 675 images |
|---|---|---|---|---|---|
| Example 1 | 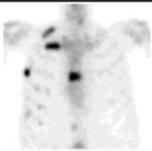 | 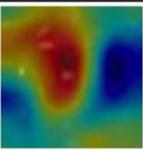 | 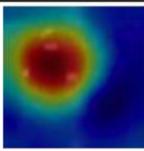 | 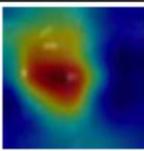 | 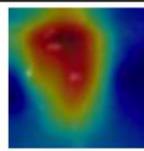 |
| Example 2 | 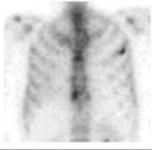 | 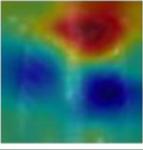 | 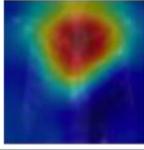 | 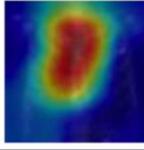 | 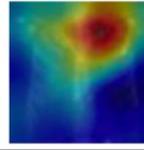 |
| Example 3 | 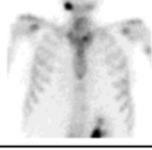 | 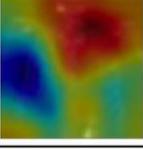 | 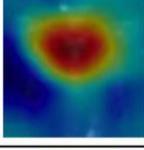 | 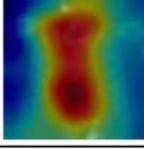 | 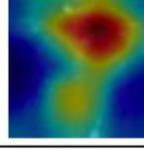 |
| Accuracy (%) | | 71.7 | 76.69 | 81.6 | 86.5 |